\documentclass[letterpaper, 10pt, conference]{ieeeconf}      
\IEEEoverridecommandlockouts                              %

\usepackage[utf8]{inputenc}
\usepackage{graphicx}
\usepackage[acronym]{glossaries}
\usepackage{float}
\usepackage{comment}
\usepackage{amssymb}
\title{\LARGE\bf Cycle Life Prediction for Lithium-ion Batteries:\\ Machine Learning and More} %

\author{Joachim Schaeffer$^{1,\dag}$, Giacomo Galuppini$^{2,\dag}$, Jinwook Rhyu$^3$, Patrick A. Asinger$^4$,\\ Robin Droop$^5$, Rolf Findeisen$^6$, and Richard D. Braatz$^{7,*}$, \textit{IEEE Fellow}
\thanks{$^1$ Technische Universität Darmstadt, Darmstadt, Germany.\newline
        {\tt\small joachim.schaeffer@iat.tu-darmstadt.de}
        }%
\thanks{$^2$ University of Pavia, Pavia, PV, Italy.\newline
        {\tt\small giacomo.galuppini@ unipv.it}
        }%
\thanks{$^3$ Massachusetts Institute of Technology, Cambridge, MA, USA.\newline 
        {\tt\small jrhyu@mit.edu}
        }%
\thanks{$^4$ Massachusetts Institute of Technology, Cambridge, MA, USA.\newline 
        {\tt\small pasinger@mit.edu}
        }%
\thanks{$^5$ Technische Universität München, München, Germany.\newline  
        {\tt\small r.droop@tum.de}
        }%
\thanks{$^6$ Technische Universität Darmstadt, Darmstadt, Germany.\newline 
        {\tt\small rolf.findeisen@iat.tu-darmstadt.de}
        }%
\thanks{$^7$ Massachusetts Institute of Technology, Cambridge, MA, USA.\newline 
        {\tt\small braatz@mit.edu}}%
\thanks{$^*$ Corresponding Author}%
\thanks{$^\dag$ These two authors contributed equally.}%
\thanks{\newline \textbf{Accepted for ACC 2024 \copyright 2024 IEEE}}
}

\newacronym{cv}{CV}{Cross-Validation}
\newacronym{lib}{LIB}{Lithium-Ion-Battery}
\newacronym{lfp}{LFP}{Lithium-Iron-Phosphate}
\newacronym{nmc}{NMC}{Nickel-Maganese-Cobalt}
\newacronym{nca}{NCA}{Nickel-Cobalt-Aluminium}
\newacronym{eol}{EOL}{End-Of-Life}
\newacronym{soh}{SOH}{State-Of-Health}
\newacronym{lli}{LLI}{Loss of Lithium Inventory}
\newacronym{lam}{LAM}{Loss of Active Material}
\newacronym{pet}{PET}{Porous Electrode Theory}
\newacronym{spm}{SPM}{Single Particle Model}
\newacronym{kmc}{kMC}{kinetic Monte Carlo}
\newacronym{mpet}{MPET}{Multiphase PET}
\newacronym{sei}{SEI}{Solid-Electrolyte Interphase}
\newacronym{dae}{DAE}{Differential and Algebraic Equation}
\newacronym{mcmc}{MCMC}{Markov Chain Monte Carlo}
\newacronym{pinn}{PINN}{Physics-Informed Neural Network}
\newacronym{pls}{PLS}{Partial Least Squares}
\newacronym{pca}{PCA}{Principal Component Analysis}
\newacronym{soc}{SOC}{State-Of-Charge}
\newacronym{rmse}{RMSE}{Root Mean Square Error}
\newacronym{mse}{MSE}{Mean Square Error}
\newacronym{eis}{EIS}{Electrochemical Impedence Spectroscopy}
\newacronym{doc}{DoC}{Depth-of-Charge}
\newacronym{dod}{DoD}{Depth-of-Discharge}
\newacronym{en}{EN}{Elastic Net}
\newacronym{ols}{OLS}{Ordinary Least Squares}
\newacronym{rr}{RR}{Ridge Regression}
\newacronym{aape}{AAPE}{Average Absolute Percentage Error}

\begin{document}
\maketitle
\thispagestyle{empty}
\pagestyle{empty}

\begin{abstract}
Batteries are dynamic systems with complicated nonlinear aging, highly dependent on cell design, chemistry, manufacturing, and operational conditions. Prediction of battery cycle life and estimation of aging states is important to accelerate battery R\&D, testing, and to further the understanding of how batteries degrade. Beyond testing, battery management systems rely on real-time models and onboard diagnostics and prognostics for safe operation. Estimating the state of health and remaining useful life of a battery is important to optimize performance and use resources optimally.

This tutorial begins with an overview of first-principles, machine learning, and hybrid battery models. Then, a typical pipeline for the development of interpretable machine learning models is explained and showcased for cycle life prediction from laboratory testing data. We highlight the challenges of machine learning models, motivating the incorporation of physics in hybrid modeling approaches, which are needed to decipher the aging trajectory of batteries but require more data and further work on the physics of battery degradation. The tutorial closes with a discussion on generalization and further research directions.
\end{abstract}

\section{INTRODUCTION}

Energy storage is vital for the transition to a sustainable future. In particular, electrochemical energy storage devices are essential for applications that require high energy- and power density, such as electric vehicles, portable electronic devices, electric vertical takeoff and landing aircraft, grid and mobile storage, and many more. This article focuses on \glspl{lib}, currently the most prominent and widely used type of electrochemical battery. However, many of the approaches that we present in this tutorial are also applicable to batteries beyond lithium-ion, such as sodium-ion or solid-state batteries.

\Gls{lib} cells consist of two porous electrodes and a separator immersed in a liquid electrolyte \cite{yoshino2012birth}. Lithium ions deintercalate from the porous cathode, pass through the separator, and intercalate into the anode while charging. The required electric work is supplied by an external circuit. This process is reversed during discharging when the battery powers an external device. The cathode usually consists of a metal oxide, such as \gls{lfp}, \gls{nca}, or \gls{nmc}, while almost all \glspl{lib} use graphite as anode material, sometimes with a small fraction of silicon to improve the anode energy density \cite{nitta2015li}. Unfortunately, during cycling and also while resting, batteries degrade. The complex interplay of many different degradation mechanisms manifests in degradation modes, such as \gls{lli} and \gls{lam}, leading to capacity fade and power fade \cite{birkl2017degradation, Edge_OKane_al_2021}. 
The challenges of understanding the interplay of degradation mechanisms led to a rise of machine learning methods for forecasting the battery cycle life (e.g., \cite{severson2019data, herring2020beep, Sulzer2021a, schaeffer2023InterpretationNullspace}). Machine learning-based battery modeling has attracted a lot of research in the last few years. One example of successful forecasting of battery cycle life in research is the optimization of battery fast charging \cite{attia2020closed}. However, care must be taken regarding data leakage \cite{geslin2023selecting}. Furthermore, many machine learning models for \glspl{lib} suffer from limited generalizability due to a lack of openly available training data \cite{Ward2022, Dubarry2020}. Many different cell chemistries \cite{preger2020degradation} and diverse operational conditions in real-world battery usage make it impossible to generate enough datasets to encapsulate all conditions that can occur in operation. Predicting cycle life accurately is important; however, predicting only the capacity is not enough for making informed decisions in R\&D and for second life \cite{aykol2021perspective}. Advancing the understanding of degradation, for example, by analyzing degradation modes from diagnostic or cycling data, is needed \cite{Vlijmen2023}. %
Furthermore, obtaining optimal performance while ensuring safety is important to use resources optimally and to safeguard life and property \cite{attia2020closed,liu2018materials,galuppini2023efficient,galuppini2024efficient}.
For this task, the battery management system estimates \gls{soc}, \gls{soh}, and ideally also \gls{eol}, which can be framed in the context of model-based diagnostics and prognostics \cite{04ReviewPersDynamicModelsKrewer2018}.

\begin{figure*}[!tb]
\centering 
\includegraphics[width=0.92\textwidth]{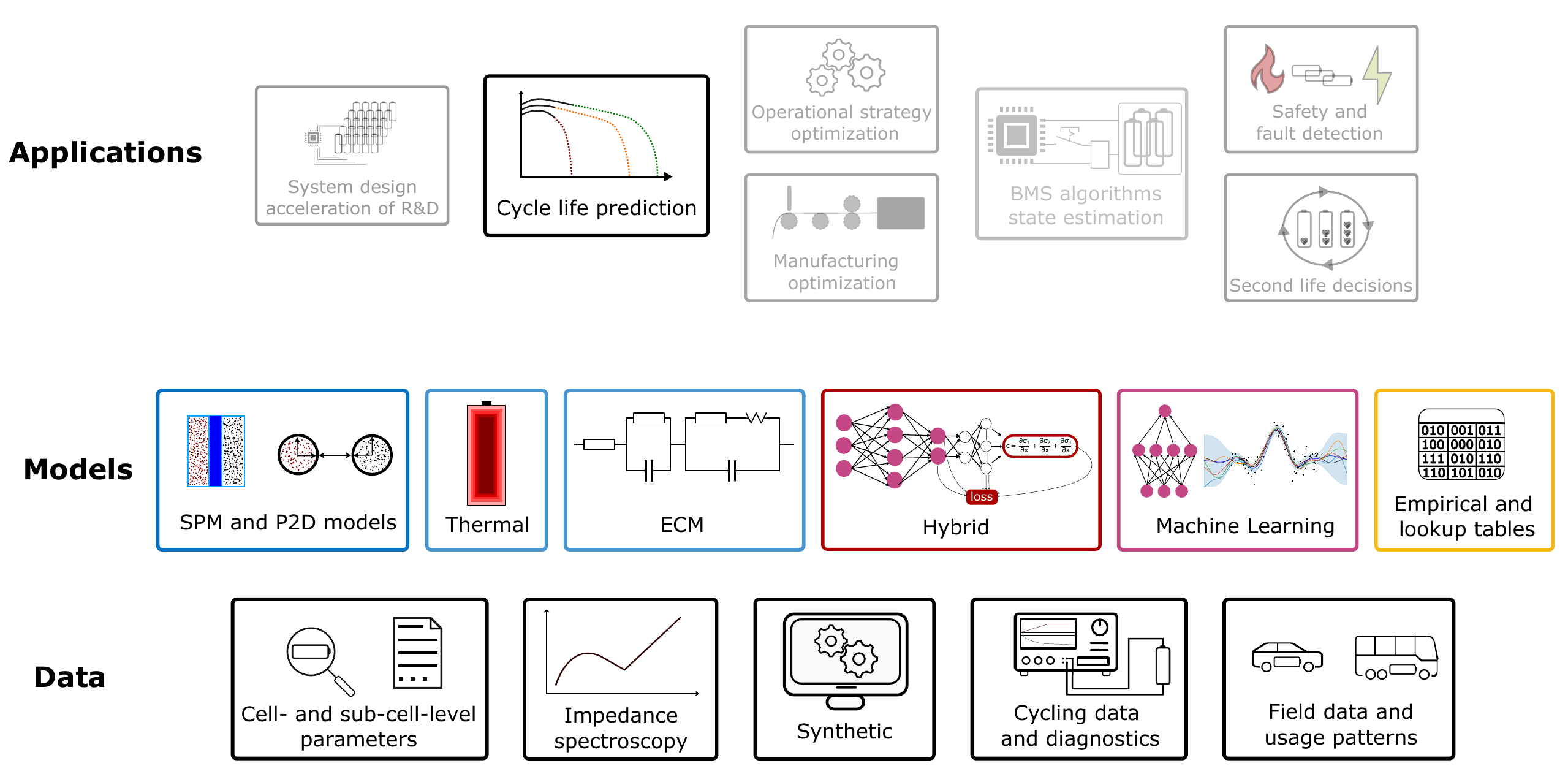}

\vspace{-0.2cm}

\caption{Data, model, and application overview for battery system modeling.} %
\label{fig:modeling_overview}	
\end{figure*}

This tutorial is structured as follows. 
The next section gives an overview of state-of-the-art first-principles, machine learning, and hybrid battery modeling approaches (middle layer, Fig.\,\ref{fig:modeling_overview}). Subsequently, 
battery cycle life prediction is showcased (top layer, Fig.\,\ref{fig:modeling_overview}), highlighting recent improvements and limitations motivating hybrid models. The last part of the tutorial discusses the importance of diagnostic data to understand degradation further and improve generalizability. %

\section{Modeling}
\textbf{First-Principles Models} describe short-term battery dynamics by using laws for reaction, diffusion, and conduction processes such as Butler-Volmer kinetics \cite{butler1924studies} and coupled ion-electron transfer kinetics \cite{fraggedakis2021theory}, yielding well-established electrochemical models (e.g., \gls{pet} \cite{newman1975porous}, \gls{spm}, \gls{kmc} model \cite{roder2017multi}, and \gls{mpet} \cite{smith2017multiphase}).
Despite their complexity, with resulting models consisting of hundreds of \glspl{dae}, several efficient implementations are nowadays available to cope with strict computing time requirements \cite{berliner2021methods}. However, first-principles modeling of degradation mechanisms and their interactions is a very active research area, with a focus on thermal effects, \gls{sei} layer growth and cracking, lithium plating, loss of active material, etc. \cite{birkl2017degradation,pinson2012theory,reniers2019review,gao2021interplay,jana2022physics}. Such physics-based models can provide a mechanistic framework (e.g., \cite{al2023physics}, \cite{downey2019physics}) to capture the influence of degradation physics on electrochemical behavior and provide good extrapolation capability, as well as transferability to a wide range of chemistries \cite{o2022modeldegradation}. However, their design is rather challenging, as it may involve degradation modes that are not yet well understood. %
A further challenge is represented by parameter identifiability issues, which may be due to the structure of the model itself, as well as by the informativity of the experimental dataset available for parameter identification \cite{berliner2021nonlinear, galuppini2023nonlinear, pozzato2023combining}.

\textbf{Machine Learning Models} gained popularity in recent years for \glspl{lib} because there is no need to deal with complex physics and challenges related to parameter identifiability.  %
For example, Severson et al.\ \cite{severson2019data} proposed a machine learning approach for predicting cycle life by combining feature engineering with a regularized linear model. %
Particularly, the features based on $\Delta \mathbf{Q}$ (i.e., the difference of discharge capacity between two cycles over voltage, Fig.\,\ref{fig:results}a) showed excellent predictive performance on a large dataset of 124 commercial \gls{lfp} cells under controlled cycling conditions. This approach demonstrated the value of integrating domain knowledge into the design of features for \glspl{lib}.
Building on \cite{severson2019data}, the Battery Evaluation and Early Prediction (BEEP) open-source Python library generates a structured feature matrix based on the provided cycling data \cite{herring2020beep}. Alternatively, the fused LASSO, an interpretable linear model, predicts the cycle life directly from $\Delta \mathbf{Q}$ \cite{schaeffer2023InterpretationNullspace}. A different approach is to use histogram-based features (e.g., \cite{greenbank2021automated}) to extend the prediction of cycle life to the capacity fade trajectory, which is important for prognostics. 
Machine learning approaches have also successfully been applied to \gls{eis} data \cite{Schaeffer_2023, jones2022impedance}. Jones et al. \cite{jones2022impedance} developed a capacity forecasting model based on \gls{eis} data that can be applied to cells with different cycling histories. %
Furthermore, several studies have also used acoustic data as a non-invasive method. Stress waves can capture physical phenomena such as bubble generation, electrode expansion, and crack formations \cite{zhang2021acoustic}, which affect degradation and performance. Although data-driven approaches can result in accurate and rapid predictions, training purely data-driven models may require a huge amount of data, and the resulting models may yield little physical insight into degradation and often struggle with generalization beyond the intended application or cell type.

\textbf{Hybrid Models} combine mechanistic and machine learning models, enabling physical insights while keeping the flexibility of machine learning models \cite{aykol2021perspective}.
As an example, the work of Ramadesigan et al. \cite{ramadesigan2011parameter} estimated the cycle life by using a data-driven model to predict the evolution behaviors of effective transport and kinetic parameters where the values at each cycle were estimated from a \gls{pet} model. In this case, the short-term battery dynamics are still governed by first-principles laws, while the long-term dynamics are described by a data-driven model. While promising, %
this approach may suffer from practical identifiability issues during parameter estimation. Furthermore, the trend of unidentifiable parameters over the cycle number yields little insight. To overcome this issue, \cite{berliner2021nonlinear} and \cite{galuppini2023nonlinear} carry out a nonlinear identifiability analysis for \gls{pet} and \gls{mpet} models, respectively. 
Once identifiability issues are properly addressed (e.g., by using Maximum-a-Posteriori estimation), the procedure can be safely applied to design hybrid models. The updated methodology is applied in \cite{berliner2023simulating} to regress kinetic and transport coefficients of a \gls{pet} model over discharge curves of 95 \gls{nca} cells from a disassembled Tesla Model 3, all cycled at least to 80\% of nominal capacity. The hybrid model developed in this way can be used to predict the battery discharge dynamics from pristine to \gls{eol} conditions.
Another hybrid modeling approach is to use a data-driven model to learn the prediction error of a previously trained first-principles model \cite{aykol2021perspective}, increasing the flexibility and prediction accuracy of the first-principles model and improving generalizability and physical insights over a purely data-driven model.
A different approach is to constrain a data-driven model to fulfill specific first-principles laws. The methodology proposed in \cite{raissi2019physics} regularizes the training loss of a neural network with the desired physical laws, leading to the design of a \gls{pinn}. This approach has three advantages: (1) increased generalizability, (2) reduced overfitting, and (3) reduced amount of training data.
Last, to overcome the need for large training datasets, data-driven models can also be trained on a combination of real and synthetic data (e.g., \cite{Dubarry2020}). The latter can be generated by previously trained first-principles models. This approach to \textit{transfer learning }\cite{weiss2016survey} has been applied successfully for battery state and health estimation \cite{shen2020deep, vidal2019li, paulson2022feature}. The next section describes an application of machine learning modeling.

\section{TUTORIAL: INTERPRETABLE \& ACCURATE DATA-DRIVEN CYCLE LIFE PREDICTION}
Ideally, the first step of building any battery modeling effort should be defining the application precisely. Next, suitable models can be selected, which then determine the data that are needed. Figure \ref{fig:modeling_overview} is a schematic of this top-down selection and decision process. Many possible paths are suited for a specific modeling task. Unfortunately, openly available battery data are scarce \cite{Ward2022}, and battery cycling experiments are expensive. Consequently, research groups wanting to explore reasonable combinations of applications and models oftentimes have no other choice than using a dataset that is not optimally suited for the investigated application, risking data leakage and overly optimistic prediction accuracies \cite{geslin2023selecting}.

Here, we revisit the problem of battery cycle life prediction from early laboratory cycling data as posed in \cite{severson2019data} and restrict the main part of the tutorial to machine learning approaches. The high-level motivation of the cycle life prediction explored in \cite{severson2019data} is to shorten laboratory battery cycling tests and accelerate the %
optimization of fast charging protocols as demonstrated in \cite{attia2020closed}.

\subsection{Data Generation}
Generally, data for cycle life prediction can range from electrode level to modules and system data. With respect to the controllability of the interfering variables, the data can be classified into laboratory, synthetic, and field data \cite{sulzer2021challenge} (Fig.\,\ref{fig:modeling_overview}).
\begin{figure}[b]
\centering 
\includegraphics[width=.4\textwidth]{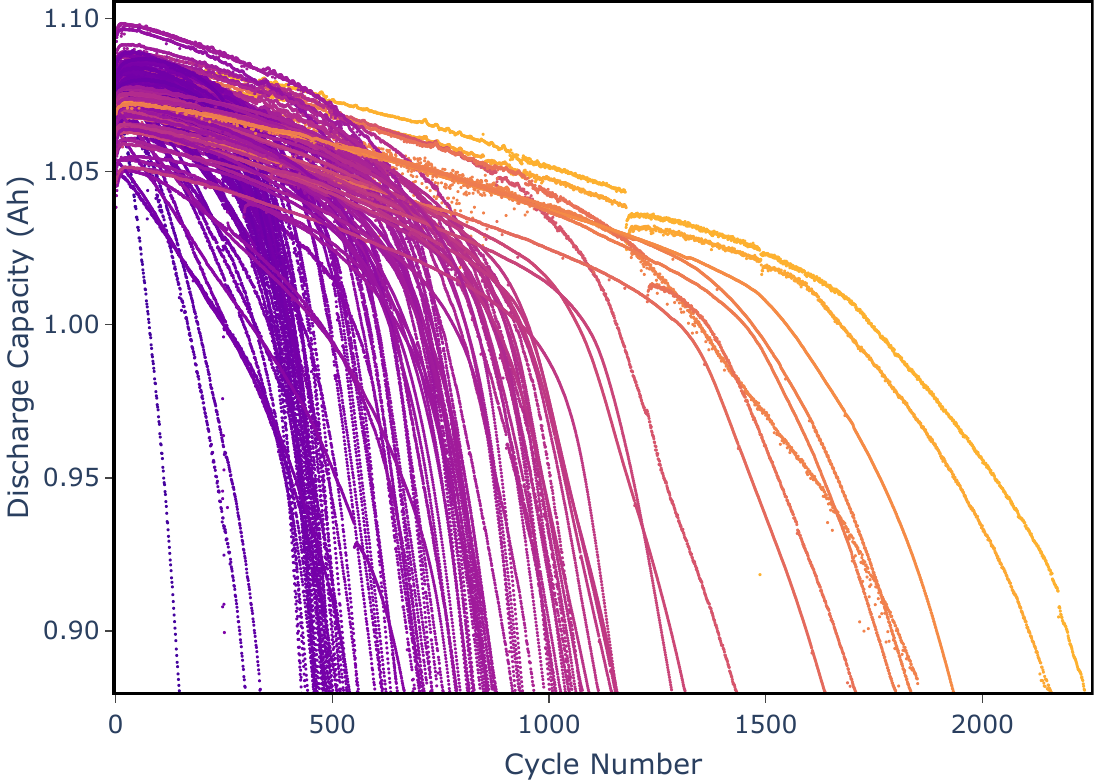}

\vspace{-0.2cm}

\caption{Discharge capacity fade over cycle number for LFP data \cite{severson2019data}.} 
\label{fig:data}	
\end{figure}
This tutorial focuses on cell-level laboratory time series cycling data. The selection of cell chemistry(-ies) and form factor(s), and the cycling experimental design (i.e., charging current, discharging current, \gls{dod} and \gls{doc}, temperature, rest times, and diagnostic cycles) are crucial design choices which depend on the research question \cite{severson2019data,attia2020closed,Vlijmen2023}. In turn, the number of cells needed depends on the range of the experimental conditions. The dataset of Severson et al. \cite{severson2019data} contains 124 cells (Fig.\,\ref{fig:data}), but \textit{only} varies the fast charging protocols between cells. All cycles are full cycles, discharge conditions are constant at 4C, the cells are cycled at a constant temperature, and all cells are cylindrical LFP cells from the same manufacturer and production batch.

\subsection{Feature Design}
Features condense and transform input data, helping the model to learn the most relevant relationships between input and output data. %
Domain knowledge is important to extract suitable and interpretable features. The $\Delta \mathbf{Q}_{100-10}$ feature (i.e., the difference of discharge capacity over voltage between cycles 100 and 10, Fig.\,\ref{fig:results}a) was developed in \cite{severson2019data}, to capture cycle-to-cycle evolutional changes in capacity and internal resistance, before significant capacity fade occurred (for more details see \cite{severson2019data}). $\Delta \mathbf{Q} \in \mathbb{R}^{124\times1000}$ is a wide matrix, where each row corresponds to a cell and each column to a discrete voltage between the voltage limits of the experiment.

Latent variable methods such as \gls{pls} or \gls{pca} construct lower dimensional representations of the data and perform regression in this latent space \cite{Schaeffer_2022}, allowing to directly use $\Delta \mathbf{Q}$ as input. Alternatively, nonlinear transformation $f\colon \mathbb{R}^p \to \mathbb{R}$, can be applied to yield scalar features. Examples are the variance and mean feature used in \cite{severson2019data}. 

A different approach is to accumulate the time spent in certain operating regions (e.g., defined by current, temperature, and \gls{soc} or voltage) and use histogram features \cite{greenbank2021automated}. This approach is also applicable to field data. However, on the LFP dataset of \cite{severson2019data}, this approach risks yielding overly optimistic prediction performances due to implicit data leakage if the charging data is used as input \cite{geslin2023selecting}. 

Feature scaling is needed for features with different units or if the absolute value of the features is very different, and we want the model to treat features as equally important a prior. Feature transformations, $g\colon \mathbb{R}\to\mathbb{R}$, such as the logarithm, root, or reciprocal function, can remove skewness in feature distributions and improve the correlation of the feature with the response variable.
Feature selection is used to improve model interpretability and generalization. For example, the Pearson correlation coefficient, or the maximum correlation coefficient for nonlinear correlations, can indicate the relevance of the features \cite{sun2021smart} and can also be used for feature selection.

\subsection{Feature Selection with Regularized Regression}
\begin{figure}[!tb]
\centering

\vspace{0.2cm}

\includegraphics[width=0.47\textwidth]{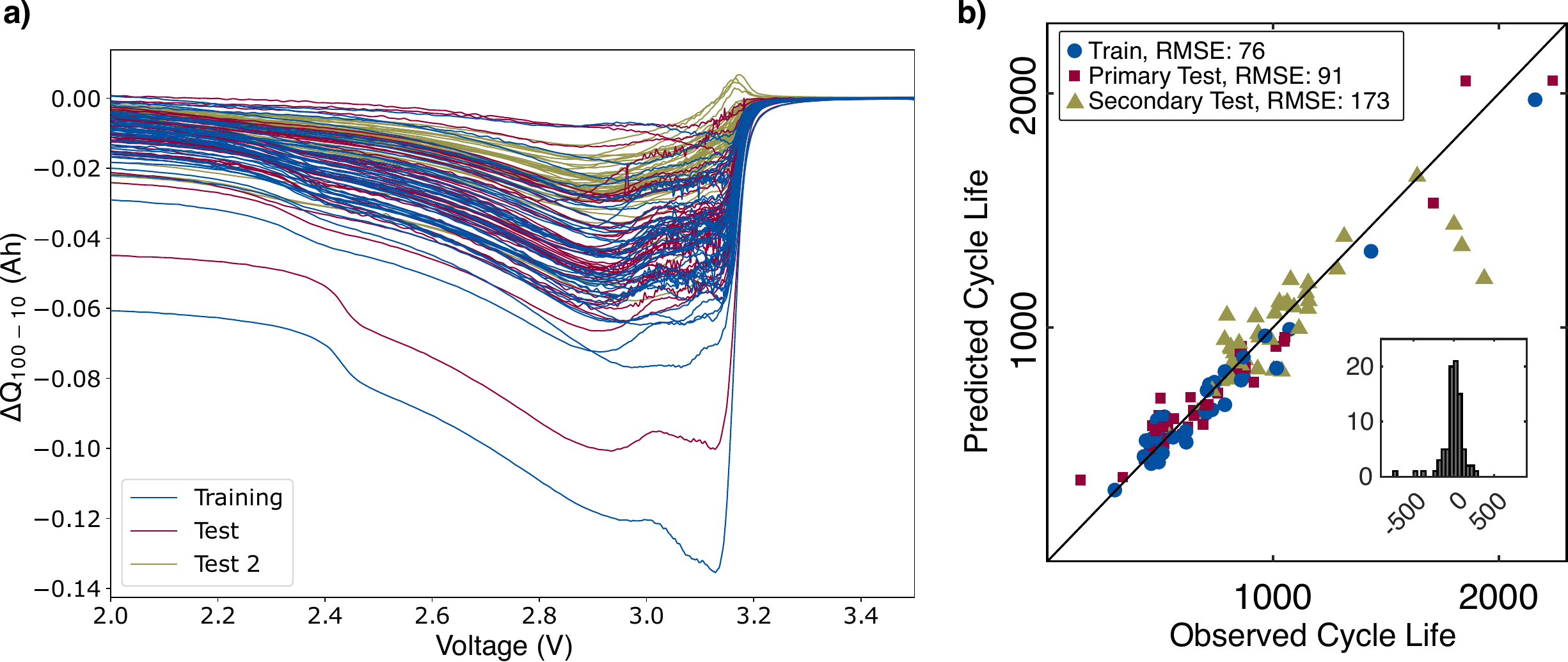}

\vspace{-0.2cm}

\caption{\textbf{a)} $\Delta \mathbf{Q}_{100-10}$, modified from \cite{schaeffer2023InterpretationNullspace}. \textbf{b)} Predictions based on an EN model using discharge features, modified from \cite{severson2019data}. Training (41 cells), primary (43 cells), and secondary (40 cells) test sets are color-coded.} 
\label{fig:results}	
\end{figure}

Consider a model of the form $\hat{\mathbf{y}}=\mathbf{\Phi}\boldsymbol{\theta}$,
where $\hat{\mathbf{y}}$ is the vector of predicted outputs (i.e., cycle life), $\mathbf{\Phi}$ is the predictor aka feature matrix, and $\boldsymbol{\theta}$ is the vector of model coefficients. The model is linear in the parameters, but its predictor matrix $\mathbf{\Phi}$ can contain arbitrary nonlinear transformations of the original input\footnote{The EN model (Fig.\,\ref{fig:results} and \cite{severson2019data}) uses scalar discharge features, with Var($\Delta \mathbf{Q}_{100-10}$) being the single most predictive feature.}. %
Assume $n$ observations are available, and consider a total of $p$ predictors (including all transformations of original inputs).
If $n\gg p$, and the identifiability condition $\textrm{rank}(\mathbf{\Phi})=p$ holds, an \gls{ols} approach can effectively estimate the model coefficients.
However, for most process data -- batteries in particular -- these conditions are usually hard to meet \cite{sun2021smart}. Typical process data are characterized by strong multicollinearity between model inputs, resulting in identifiability issues \cite{hastie2009elements}. The feature engineering approach may also introduce further collinearity as a side effect. 
Regularization is an effective way to reduce identifiability issues and perform automatic selection of relevant predictors while at the same time providing an estimate of the model coefficients \cite{hastie2009elements}.

The \gls{en} is a regularized identification algorithm that combines the advantages of \gls{rr} and LASSO  \cite{hastie2009elements}. The \gls{en} estimate of the model coefficients is
\begin{align}
\nonumber
    \hat{\boldsymbol{\theta}}^{\textrm{EN}}\!=\arg \min_{\boldsymbol{\theta}} ||\mathbf{y}-\mathbf{\Phi}\boldsymbol{\theta}||^2_2+\lambda \Big(\frac{1-\alpha}{2} ||\boldsymbol{\theta}||_2 + \alpha||\boldsymbol{\theta}||_1 \Big)
\end{align}
where $\lambda$ and $\alpha$ are hyperparameters. Specifically, the {1-norm} regularization associated with the LASSO allows feature selection by zeroing the parameters, i.e., discard predictors. The {2-norm} regularization from \gls{rr} improves feature selection by consistently pushing towards zero the parameters of all collinear features \cite{hastie2009elements}.
Rigorous \gls{cv} should be used to select $\lambda$ and $\alpha$. \Gls{cv} estimates the expected out-of-sample prediction error by splitting the training data into subsets, repeatedly training the model on all but the holdout set, and evaluating the model performance based on the holdout dataset \cite{sun2021smart,hastie2009elements}. The estimated best hyperparameter values are those resulting in the model with the lowest average prediction error on the holdout data. Finally, the model is trained on all available training data using the best hyperparameter values found. Note that battery data are often characterized by an underlying group structure due to, e.g., batched production processes or in-situ operating conditions. To avoid fitting possible system biases, \textit{grouped} \gls{cv} \cite{sun2021smart} should be used, which splits data at the \textit{group} level, ensuring that all of the data belonging to a group is assigned to the same subset.

\subsection{Model Reliability, Predictive Performance, and Intepretability}

Many metrics are available to evaluate a model's predictive performance. Commonly used examples are  \gls{mse} and \gls{rmse}, $R^2$, \gls{aape} \cite{hastie2009elements}.
The use of (possibly multiple) \textit{test} datasets is fundamental to avoid a biased evaluation of the predictive performance of the model. Moreover, this approach allows an even comparison of the performance achieved by different models (e.g., models based on different candidate predictors) \cite{hastie2009elements}. 
The analysis of model residuals $\boldsymbol{\varepsilon}=\mathbf{y}-\mathbf{\Phi}\boldsymbol{\theta}$ gives indications regarding the reliability of the model structure. Residuals resulting from a ``good" model should be zero mean, normally distributed, and independent of $\mathbf{y}$. The analysis involves both visual inspection of the residual histogram and Q-Q plots, as well as statistical tests such as the $\chi^2$ test \cite{hastie2009elements}. 
Another fundamental issue is the characterization of the uncertainty affecting both the estimated parameters and the model prediction. The former aspect, which can be addressed by, for example, estimating the variance of model parameters and confidence intervals/hyperellipsoids, gives clues about the reliability of the model. An example is the case of confidence intervals spanning over positive and negative values for a parameter. In this case, the predictor can possibly be discarded from the model, as it is not even possible to significantly know the sign of its contribution to the model prediction \cite{hastie2009elements}. The second aspect is instead important when decisions must be taken based on model predictions (e.g., timely replacement of cells in biomedical applications or design of safe charging protocols). 
A final caveat is related to model interpretation. %
The fact that any particular specific predictor allows for good predictive performance does not imply the existence of a direct, causal relation between the predictor and the target variable. Therefore, any cause-effect inference inspired by the model structure or choice of predictors should be supported by domain expertise and physics-based arguments \cite{sun2021smart}.

\subsection{Battery Cycle Life Prediction Case Studies}
Severson et al.\ \cite{severson2019data} designed features based on $\Delta \mathbf{Q}$ and subsequently used an \gls{en} for determining the regression coefficients. Figure \ref{fig:results} shows the resulting predictions on the training, primary, and secondary test data. The \gls{rmse} error is higher for the secondary test dataset; in particular, the model struggles to accurately predict the cycle life of the longer-lived cells in the secondary test dataset (Fig.\,\ref{fig:results}b). The cells in the secondary test dataset were cycled approximately ten months after the cells in the training and test set and thus experienced higher calendar aging. Furthermore, few cells with a cycle life $>1200$ cycles are contained in the training dataset.  
Recently, Schaeffer et al.\ \cite{schaeffer2023InterpretationNullspace} showcased a different approach, using the fused LASSO -- a generalization of the LASSO particularly useful for discrete measurement data over a continuous domain -- to directly learn from $\Delta \mathbf{Q}$. The resulting model is highly interpretable due to piecewise constant regression coefficients and yields similar prediction performance to the discharge model on the cells with cycle life $<1200$ cycles. %

\section{DISCUSSION ON GENERALIZATION AND CONCLUSION}
Despite many recent advances, a key challenge of machine learning models for battery cycle life prediction is the risk of data leakage, which means that models will underperform in the target application compared to their in-house results. Careful experimental planning and model design are necessary to ensure the right features are constructed for the intended application.
Furthermore, machine learning models risk overfitting to the specific experimental conditions. For example, while predictive models trained on \textit{accelerated cycle aging data} may hasten research cycles and quality control, their generalization to real-world operation would likely be poor due to their ignorance of non-cycling aging (i.e., calendar aging). Given that many battery deployments spend most of their time idle (e.g., EVs), the influence of calendar aging must be accounted for in most practical lifetime prediction applications. The exclusion of calendar aging in lifetime predictions may lead to overly optimistic lifetime projections.
Insights from machine learning models based on \textit{high C-rate} data are limited; even the most interpretable models, such as the fused LASSO, can only support musings about underlying physics and degradation reactions. 
Therefore, diagnostic cycles that quantify degradation in rich and complementary ways are necessary. For example, (1) performing slow (e.g., C/25) charge and/or discharge to obtain high fidelity pseudo-OCV data to conduct differential voltage analysis to quantify degradation modes such as \gls{lli} and cathode and anode \gls{lam}, and (2) performing pulse tests with various magnitudes and directions at multiple states of charge and temperatures to capture resistance changes. The complementary pseudo-equilibrium and dynamic characterizations help to paint a more detailed picture of cell degradation. \cite{Vlijmen2023} demonstrates this kind of experimental design and analysis and opens many new research avenues, such as hybrid models, to improve robustness, generalization, and physical understanding. Diverse cycling data combined with rich diagnostics is essential for improving the understanding of battery degradation and thus improving battery design and management.

To conclude, this tutorial showcases different battery models and recommends a top-down decision-making process, referring to the layer in Fig.\,\ref{fig:modeling_overview}, for battery modeling R\&D. The tutorial section showcased battery cycle life prediction, including recent advances on the dataset in \cite{severson2019data}. Furthermore, we discussed limitations and directions for future research to improve the understanding of degradation and, in turn, build better batteries. 

\bibliography{Bibliog.bib}
\bibliographystyle{IEEEtran}

\end{document}